\documentclass[runningheads]{llncs}
\usepackage[T1]{fontenc}
\usepackage{amsmath,amssymb,amsfonts}
\usepackage{graphicx}
\usepackage{pgfplots}
\usepackage{courier}

 \usepackage{epstopdf}
\usepackage{textcomp}

\usepackage{algorithm}
\usepackage{pdfpages}
\usepackage{subcaption}
\usepackage{algpseudocode}
\usepackage{xcolor}
\usepackage{multirow}
\usepackage{comment}
\usepackage{adjustbox}
\usepackage{mathtools}
\usepackage{url}
\usepackage[utf8]{inputenc}

\pgfplotsset{compat=1.18}

\begin{document}
%
\title{TapTree: Process-Tree based Host Behavior Modeling and Threat Detection Framework via Sequential Pattern Mining}
%
%

\author{Mohammad Mamun \and Scott Buffett  }
\authorrunning{M. Mamun \& S.Buffett}
%
\institute{National Research Council Canada, Fredericton, NB, Canada 
\email{\{Mohammad.Mamun,Scott.Buffett\}@nrc-cnrc.gc.ca}}

\maketitle              
\begin{abstract}

Host behaviour modelling is widely deployed in today's corporate environments to aid in the detection and analysis of cyber attacks. Audit logs containing system-level events are frequently used for behavior modeling as they can provide detailed insight into cyber-threat occurrences.
However, mapping low-level system events in audit logs to high-level behaviors has been a major challenge in identifying host contextual behavior for the purpose of detecting  potential cyber threats. 
Relying on domain expert knowledge may limit its practical implementation. 
This paper presents TapTree, an automated process-tree based  technique to extract host behavior by compiling system events' semantic information.
After extracting behaviors as system generated process trees, TapTree integrates event semantics as a representation of behaviors.
To further reduce pattern matching workloads for the analyst, TapTree aggregates semantically equivalent patterns and optimizes representative behaviors.
In our evaluation against a recent benchmark audit log dataset (DARPA OpTC), TapTree employs tree pattern queries and sequential pattern mining techniques to deduce the semantics of connected system events, achieving high accuracy for behavior abstraction and then Advanced Persistent Threat (APT) attack detection. Moreover, we illustrate how to update the baseline model gradually online, allowing it to adapt to new log patterns over time. 
\keywords{Process tree \and Behavioral Anomaly Detection \and sequential pattern mining \and APT detection}
\end{abstract}

\section{Introduction}



Since modern information systems have become critical and essential components of contemporary businesses and organisations, insider threat detection is becoming a rapidly growing topic of study in the cybersecurity domain. An emerging cyberattack, known as APT, poses a huge threat to these information systems, first by breaching hosts inside a target system and then stealthily infiltrating additional hosts through the internal network to steal sensitive information. Since attackers often sabotage legitimate services executing on endpoints, it is critical to detect malicious behaviour on endpoint computers promptly and efficiently following a breach, prior to major harm being caused. Several recent studies demonstrate that malicious behaviour can be detected by leveraging patterns of benign behavior against other, seemingly benign actions that, when combined, signal something potentially more destructive \cite{log2vec2019,Mamun2021,Du2017,APTmodel}. 

Unfortunately, the volume of log events produced by a typical host is huge. For instance, a single desktop computer, let alone servers in large enterprise network, can generate over a million events each day \cite{lee2013loggc}. 
Processing massive amounts of audit log events and filtering out irrelevant system events in order to recognize representative host behavior requires a tedious manual effort \cite{liu2018towards}. Existing solutions to this problem include techniques such as tag propagation \cite{hossain2017sleuth} and graph matching \cite{zong2015behavior,log2vec2019,han2020unicorn}, that mostly rely on domain expert knowledge or on a knowledge store of expert-defined rules \cite{zeng2021watson}.
To address this issue, our objective is to develop an efficient method for extracting representative behavior (i.e. \cite{zeng2021watson,Mamun2019}) for cyber analyst investigation. More precisely, we automate the extraction of host behavior using procedural task analysis on system log events and then aggregate semantically related tasks to construct baseline host behavior. Due to the fact that recurring or similar tasks have been aggregated together, TapTree can significantly reduce the number of events to analyze.

Existing anomaly detection approaches convert user operations into sequences to analyze sequential relationship between log entries, and then employ sequence processing techniques, such as deep learning \cite{Du2017,Zhang2016,PanPan2019,Mamun2021,LogNADS2021}, natural language processing \cite{Young2019}, to learn from previous events and predict the next event. These methods at the log-entry level model user behaviour and indicate discrepancies as anomalies. 
However, this approach is oblivious to other relationships. For example, a user's daily activity is \textit{relatively} regular over time in terms of the logical relationship among periods \cite{log2vec2019}. Moreover, event logs may also be generated concurrently by many threads, aliases, or tasks \cite{Mamun2021}. If this relationship in the log is disregarded, prediction methods based on continuous logs may suffer a loss of reliability.

We construct a baseline behavior model and assess its ability to correctly detect malicious behaviour on a recently released APT attack dataset (DARPA OpTC public dataset ~\cite{OpTC2020}). Evaluation results on 14 randomly selected hosts from the OpTC dataset show that TapTree recognizes targeted host behavior with accuracy over 99\% with false positive rate (FPR) of less than 0.8\% when using tree pattern queries, for a given candidate partial match threshold. We show that this threshold can then be adjusted to cast a wider net and achieve a perfect 100\% recall on malicious behaviour detection, while still keeping the FPR relatively low at 2.9\%. For the sequential pattern-based analysis, we show that FPR can be further reduced to below 0.1\%, while maintaining high accuracy (>99.9\%) and recall (67\%), whereas recall can be improved to 86\% while still maintaining FPR < 1\%. Moreover, we quantify the proportion of process trees that are reduced in number as a result of similar pattern aggregation. Our results demonstrate that TapTree can reduce the number of process-trees from raw audit logs by 98\% percent after aggregation and thus substantially reduce the analysis overhead associated with abnormal behavior investigation. Our major contributions are summarized:

\begin{itemize}
    \item We present TapTree, a process tree-based host behavior modeling. TapTree automatically encapsulates host contextual behaviors from raw audit log events using system generated process-tree. To our knowledge, this is the first approach to host behavioral abstraction that utilizes system process trees to aggregate semantically equivalent patterns.
    
    \item To reduce analysis overhead for the analyst and enable efficient detection, TapTree considers noise reduction, optimized tree growth such as forward pruning and aggregation of similar behaviors.
    
    \item As part of validation of proposed behavioral abstraction model, we conduct a systematic evaluation by abstracting benign behavior in a given context e.g. APT attack against Darpa-OpTC dataset.
    We propose the use of sub-tree pattern and sequential pattern queries to detect discriminatory behaviour and identify insider threats automatically. 
    Experimental results using Darpa-OpTC data demonstrate that TapTree's baseline behavior is effective against both benign and malicious behaviors.
    In terms of mining speed, TapTree outperforms baseline generation model without aggregation by two orders of magnitude.

\end{itemize}

\subsection{Analyzing the problem} 
Two characteristics must be satisfied for behavioral model to be deemed effective: (i) behavioral distinctiveness to accurately represent the host behavior and (ii) behavioral consistency to identify deviant behaviors \cite{mazzawi2017anomaly}. We study ways to satisfy these requirements while decreasing or eliminating the bulk of false positive occurrences.

To address this issue, our approach is to construct heterogeneous temporal process trees from homogeneous system events representing the target behavior and then use the forest (group of trees) to build the host model.
Note that the number of system events might be enormous, highly interconnected, and noisy. For instance, the installation of a single package during an APT campaign 
may create over 50 thousand system log events \cite{zeng2021watson}. In addition, the number of log entries containing information about suspicious/malicious activity is quite small. Rather than using all trees, we select the most discriminating patterns in the forest by grouping similar/redundant actions to aid analysts' analysis (e.g. if a behavior is a subset of another, it gets merged).
Such representative behavior-specific forest (with modest number of trees) is simpler to interpret, faster to match, and easier to maintain.

\begin{figure*}[h]

    \caption{Sample task trees (process name, depth, occurrence-edge)} 
    \label{Sample Task tree}
    \hspace*{\fill}   
    \begin{subfigure}{0.50\textwidth}
        \centering
        \includegraphics[width=\textwidth,height=6cm,keepaspectratio]{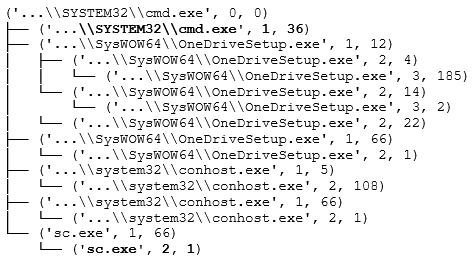}
        \caption{Benign}
        \label{Benign }
    \end{subfigure}%
    \begin{subfigure}{0.50\textwidth}
    \centering
        \includegraphics[width=\textwidth,height=6cm,keepaspectratio]{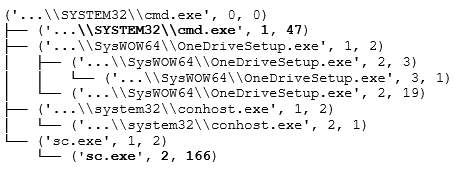}
        \caption{Malicious}
        \label{Malicious}
    \end{subfigure}%

\end{figure*}

A representative pattern should be frequent in the intended representative behavior and rare in deviant behavior. For example, a system administrator often logs into servers and does certain tasks that a human resources professional performs seldom. A sample benign/malicious tree pattern from Darpa-OpTC is presented in Figure~\ref{Sample Task tree} containing the number of edges occurrences in the pattern.

Detection algorithms are primarily based on two pattern matching techniques: 1) a tree search method for pattern matching (Section \ref{Tree pattern queries}) and 2) a sequential pattern mining classification algorithm (Section \ref{Sequential pattern classification}) on the sequence generated from the Temporal tree set as discussed in Section \ref{Tasks Forest Generation}.

\begin{figure*}[h]
\centering
   
    \hspace*{\fill}   
    
        \includegraphics[width=\textwidth]{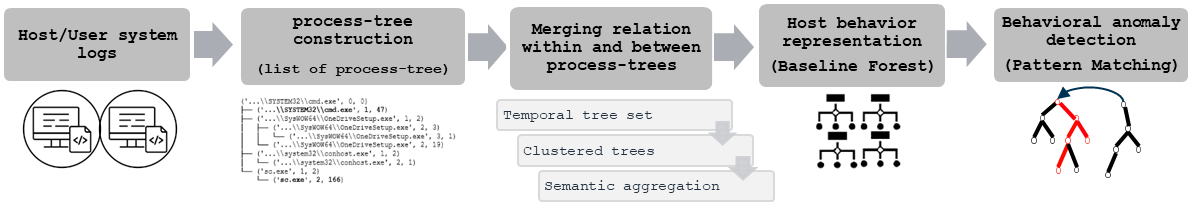}
        
         \caption{TapTree pipeline for behavioral anomaly detection} 
         \label{Architecture}

\end{figure*}

\section{Overview}


\subsection{Assumption} 
TapTree is designed to work with complex event data structures that are both hierarchical and sequential in nature (filiation-relationship of events). 
We presume that behaviors are audited at the kernel level and their activities are logged in system-call audit logs. The integrity and security of the underlying audit log monitoring platform (SIEM security) are beyond the scope of this study and are thus considered to be a component of the trusted computing base.

Rather than focusing on a single user's sessions \cite{zeng2021watson} or a single user's one day data \cite{cochrane2021sk}, we target multiple users' whole dataset (7 days data) for behavior modeling and validation. Results show that our approach is broadly applicable across hosts, days/sessions behavior.

A naive way to obtain the semantic representation of a host behavior is to gather up individual tasks (collection of process-trees) derived from its component events. This approach, however, may overlook the \textit{relative weights of relationships} between events (edges in the tree) and \textit{noisy relations} in the representative behavior.  

A process-tree is a collection of relationships (edges) between low-level operations such as process-creation, file-opening, etc., triggered by user activity. We assume that benign tasks (typical user behavior) have a strong correlation. However, there may be fewer connections between benign and abnormal operations. While these operations also mirror user behavior, not all of them contribute to the semantics of host behavior. In light of these observations, TapTree identifies the relationships that are frequent within a task (process-tree) and across tasks/behaviors. This approach provides a higher discriminative weight to the relationships in the process-tree that are less prevalent.


\section{System Design}
\label{Baseline model}


TapTree is a host behavior modeling and threat detection system. It is comprised of three main components: process-tree construction, representative behavior generation, and behavioral stability evaluation (e.g. anomaly detection). Figure~\ref{Architecture} depicts TapTree's detailed approach that takes system audit logs as input data, generate temporal process-trees as individual tasks/behaviors, aggregates/abstracts behavior semantics to output representative behaviors. 

\subsection{Process-tree construction}

Hierarchical structure of process tree derived from an audit log reflects causal relationships between running processes of a computer system. Besides providing a holistic view of the system process life cycle, this property of the system process tree offers valuable contextual information about an event's proximity continually evolving over time \cite{Kent2015}. 
The first component of TapTree is a process-chain based heuristic technique for mapping relationships between log entries that reflect hosts' behavior across several streams, such as file operation, authentication, flow, etc., into a task-process-tree.

TapTree primarily considers three types of relationships for generating process-trees: 1) the filiation relationship that forms a hierarchy across all running operating system processes, 2) sequential relationship between traces and 3) the logical relationship among tasks.

A task-process-tree (see Fig. \ref{Sample Task tree}) is a temporal semantic behavior  tree represented by a tuple $T := (V, E, R)$:

\begin{itemize}
    \item $V$ is a set of nodes where $v \in V$ represents a path to the program (e.g. \texttt{\textbackslash \textbackslash System32\textbackslash \textbackslash conhost.exe}) that initiates an event (e.g. \texttt{Process-creation})
    
    \item $E \subset V \times V \times R $ is a set of directed edges where $e =(u,v,r) \in E$ denotes a chronologically ordered relationship between executing programs.
    
    \item $R$ is a set of possible occurrences between nodes $V$, where $r\in R$ is a positive integer. Therefore, each $e \in E$ is assigned a weight $w(u,v): \mathbb{R}^+$ that implies the frequency/occurrence of the two program $(u \rightarrow v)$ invoking each other.      
    
\end{itemize}

The process-tree data associated with a target behavior/task is used to construct a program-path tree, as these behaviors are typically executed by a single thread.
This work focuses on the program-paths since our empirical findings show that using program-path instead of raw events is quite effective at abstracting host behaviors. Additionally, it offers significant computational benefits over more complex provenance/knowledge graph models.



%







\subsection{Fusion of host behaviors}
\label{Baseline model comparison}
A behavior instance, such as a process-tree in our case, consists of a series of events connected semantically. Program-path identifies the path to the program initiated by an event. A fine-grained associations between these behavior instances can provide high-level abstraction for generating effective behavioral model.

TapTree consolidates behavior instances using two widely used approaches for behavioral abstraction--- path-based approach \cite{wang2020you,Mamun2021,Du2017} by splitting up process-trees into paths  and contextual-representation based approach \cite{zeng2021watson} by extracting sub-tree as an instance of a behavior. The following sections cover TapTree's approach to behavior consolidation.

\subsubsection{Temporal tree set generation} 
\label{Tasks Forest Generation}

A temporal tree set is a collection of unique task process-trees where the trees with the same number of elements and relations/edges are merged. 
Because the trees are weighted, the weight of an edge between two or more similar trees is equal to the sum of the weights of its edges.
As the trees are weighted, the edge's weight is equal to the maximum of its edges' weights.

Formally, a temporal tree set $F \subseteq \mathbf{T}$ of task trees is a set of $(n \geq 0)$ disjoint weighted directed trees such that,
\begin{itemize}
    \item For all $ P=(V_P,E_P, R_P), Q=(
V_Q,E_Q,R_Q) \in F$,  $V_P \neq V_Q$, $E_P \neq E_Q$ ,$R_P \neq R_Q$,
\item For all $ r \in R$ where $V_P = V_Q$ and  $E_P = E_Q $, $p \in P$ and $q \in Q$
\\ $r = Max(\mathbf{w}_p(e_p), w_q(e_q))$. 
\end{itemize}

 

\subsubsection{Clustering trees}

This method consolidates the relationships within a tree in order to avoid repeating patterns. Duplicate relations/edges at the leaf level of the tree are merged in this stage.

A clustering of the leaves of the tree $T$ can be defined by cutting a subset of edges $C \subseteq E$. One method for achieving this is to solve the \emph{max-diameter min-cut partitioning problem} \cite{Balaban2019}. 
We define a partition level $ \{ L_1, L_2, \dots, L_N\}$ of $L$ to be an admissible clustering if it can be obtained by removing some edge set $C$ from $E$ and assigning leaves of each of the resulting connected components to a set $L_i ($ where $N \leq |C| + 1)$.

Let $T=(V,E,R)$ be a directed tree containing two edges $e_1= (u_1,v_1,r_1)$  and $e_2 = (u_2,v_2,r_2)$ with  $u_1 \neq v_1, u_2 \neq v_2, u_1 = u_2, v_1 = v_2$ where \{$v_1, v_2$\} are leaf nodes. Merging $e_1$ and $e_2$ results a new tree $T^\prime=(V^\prime,E^\prime,R^\prime)$, where $V^\prime = (V\setminus \{u_2, v_2\})$, 
$E^\prime = (E\setminus \{e_2\})$, $r^\prime \in R^\prime = r,$ or $ Max(\mathbf{w}(e_1), w(e_2))$ if $r_1 \neq r_2$.

\begin{table}[ht]
\centering
\footnotesize
\caption{TapTree baseline model generation and matching efficiency}
\begin{tabular}{|c|ccc|}
\hline
Method                                                      & \#Trees   & \begin{tabular}[c]{@{}c@{}}Baseline Generation\\ (in \textit{s})\end{tabular} & \begin{tabular}[c]{@{}c@{}}Pattern Matching \\ (in \textit{ms})\end{tabular} \\ \hline

\begin{tabular}[c]{@{}c@{}}Temporal Tree set\end{tabular}       &     3501                                                          &   135.51          &  $-$                                                                  \\

\begin{tabular}[c]{@{}c@{}}Clustered Trees\end{tabular}     & 2372                                                              &    163.25         & 65.3992                                                            \\

\begin{tabular}[c]{@{}c@{}}Semantic Aggregation \end{tabular} & 
  36  &   4301.83              & 47.0608                                                            \\

\hline
\end{tabular}
\label{Table:Genreation}

\end{table}

\subsubsection{Semantic aggregation}



After redundant instance aggregation, we deduce the semantics of a behavior instances naturally by combining trees derived from clustered trees. 
Identifying a pattern, whether it is a new one to aggregate or a previously discovered one, can help in averting instances of repetitive behavior.

Recall, a naive way to obtaining the semantic representation of a behaviour instance is to consider all the trees derived from the events. However, this approach may work only if the baseline semantics of behavior (temporal tree set) is decently small or it does not need updating over time. In practice, this technique is not efficient from the view point of detection (matching) for a large enterprise system where thousands of flow of events need to be examined in a certain period.      
Additionally, this assumption is frequently incorrect due to the way tree relations are weighted differently to represent the semantics of behaviour and the effect of noisy events.




\noindent \emph{Induced  subtree}: Given a tree pair $(T_1,T_2)$ where $T_1 := (V_1, E_1, R_1), T_2:= (V_2, E_2, R_2) $, we say $T_2$ is an induced subtree of $T_1$ denoted by $T_2  \preceq  T_1$, if and only if,

\begin{enumerate}

    \item $V_2  \subseteq V_1$ and  $E_2  \subseteq E_1$,
    \item Filiation relationships in $T_2$ must be preserved in $T_1$. That is, parent-child relations for all $ e= (u,v) \in T_2$ is identical to that of $T_1$, 
    \item The left-to-right ordering of siblings in $T_2$ must be a subordering of the associated nodes in $T_1$.

\end{enumerate}

\noindent \textit{Growing baseline trees}: Using consecutive growth options (forward, backward, and inward) as described in \cite{zong2015behavior} for searching a given behavior pattern against baseline patterns ensures a complete and non-repetitive search in the pattern space. In this manner, behavior pattern trees are iteratively merged if they are not \textit{induced subtrees} of the baseline trees in order to construct a baseline behavior model.

Let $T_a = (V_a, E_a, R_a)$ and $T_b = (V_b,E_b, R_b)$ be two directed trees.
A merging of two trees $(T_a \cup  T_b)$ results a new tree $T_{ab} = (V_{ab}, E_{ab})$ such that $E_{ab} = E_a \bigcup E_b$ and $V_{ab} = V_a \bigcup V_b$ that satisfies $V_{a1} \in V_b$  or $V_{b1} \in V_a$ where $V_{a1}$ and $V_{b1}$ are the roots of $T_a$ and $T_b$ respectively.

Table \ref{Table:Genreation} outlines a comparative analysis of the aforementioned methods for behavior consolidation in relation to baseline construction. We present the volume of behavior patterns and the execution time (in seconds) required to generate the pattern in each phase of the baseline generation model in the number of trees and baseline generation column of the table.


\subsection{Behavioral anomaly detection}

Behavioral anomaly detection (BAD) is expected to effectively resolve a variety of security issues by detecting deviations from a host's normal behavioral patterns. BAD enables the monitoring of applications for malicious behavior (e.g. intrusion, compromise detection), thereby improving protection against \textit{Zero Day} attacks. 
Host behavior abstraction model discussed in the previous section can be used for behavioral anomaly detection such as APT attacks. Given the behavior representation for any host or server we can utilize  (1) unsupervised model/ one-class classifier (tree pattern matching) or (2) supervised model/ binary classifier (sequential pattern matching) to identify evidence of anomalous behavioral events.

The tree pattern matching algorithm compares a sequence of operations to a baseline model in order to determine whether a task is abnormal. The tree search method allows for a trade-off between recall and the false positive rate in detection. Note that an exhaustive search of the tree will always return the prototype that is closest to the input vector. However, alternative search methods can be used to determine a task that is a close match to the baseline but not necessarily an exact match.

Sequential pattern-based analysis works specifically on a set of event {\em traces} (i.e. sequences) extracted from the task trees, and identifies common temporal patterns that reside in those sequences. This can be used to establish a model of baseline activity, against which new activity can be measured to determine the likelihood that the new activity appears as expected and is not malicious, or to construct a classification model on labeled data in the case that sufficient samples of malicious activity can be obtained.

In the following, we discuss tree pattern queries and sequential pattern queries in detail that were used to evaluate TapTree's efficacy in identifying behavioral abnormalities.

\subsubsection{Tree Pattern Queries.}
\label{Tree pattern queries}

We conduct a systematic study of tree matching algorithms that determine the likelihood of a pattern occurring by performing a recursive comparison on each node of the tree. When a mismatch is detected, the comparison procedure is terminated.

Typically, queries on trees are executed using one of two classic graph traversal strategies: breadth-first search (BFS) or depth-first search (DFS). We use a modified DFS graph-querying algorithm for tree pattern queries. DFS can expand one intermediate result at a time, starting from the first variable in the pattern and continuing to the
next ones until the whole pattern is matched. DFS can expand a single intermediate result at a time, beginning with the first variable in the pattern and progressing through the remaining variables until the entire pattern is matched.


Let $(T_i,P)$ be the baseline trees (target host behavior model) $T$ and a pattern tree $P$ pair where children of all nodes are labelled and ordered. $P$ matches at node $t$ if there is a $1-1$ mapping from nodes of $P$ to $T$ such that:  1) root of $P$, $R_P \leftrightarrow{} t$ and 2) if  $\exists (i\in P \leftrightarrow{} j \in T) $, all the children follows. Let $\lambda_v$ be the path from $R_P$ to $v$. $v$ matches $T$ at node $u \in T$ if $\lambda_v$ matches $T$ at $u$.  \\

\noindent \textit{Exact Match:} In this method, the pattern tree $P$ must be matched \textit{exactly} with any of the trees in the baseline patterns with respect to node label, inheritance, and order relationship. An exact match of a pattern $P$ into a baseline tree $T$ is a mapping $\mathcal{F}_{exact}: P \xrightarrow{} T$ for each nodes of $P$ that satisfies: 

\begin{itemize}
    \item  For each $u \in P,$ $label(u) = label(\mathcal{F}(u))$
    \item  If $\exists u_i \rightarrow u_j \in P$ then $\mathcal{F}(u_i)$ is a parent of $\mathcal{F}(u_j) \in T$. If $u_i \Rightarrow u_j \in P$, $\mathcal{F}(u_j)$  is a descendant of $\mathcal{F}(u_i) \in T$  
    \item  For any edge $e: u_i \Rightarrow u_j \in P$ where $label(u_i) = label(\mathcal{F}(u_i))$  and $label(u_j) = label(\mathcal{F}(u_j))$, $ e(c) \leq \mathcal{F}(e(c))$ where $c$ is the frequency count of the relation such as $u_i \Rightarrow u_j$ 
    \item For any $ (u_i, u_j) \in P$ if $u_i$ is to the right of $u_j$, $\mathcal{F}(u_i)$ is to the right of $\mathcal{F}(u_j)$.
    
\end{itemize}

\noindent \textit{Partial Match:} In this method, the pattern tree $P$ must be matched partially with any of the trees in the baseline model such that the root element and all elements connected directly and indirectly to the root are matched with respect to node label, inheritance and order relationship to the  baseline tree. A partial match pattern $P$ into a baseline tree $T$ is a mapping $\mathcal{F}_{partial}: P \xrightarrow{} T$ that returns $R$ that satisfies:

\begin{itemize}
    \item Let $\exists R$
    \item  If $\exists u_i \rightarrow u_j \in P$ then $\mathcal{F}(u_i)$ is a parent of $\mathcal{F}(u_j) \in T$. If $u_i \Rightarrow u_j \in P$, $\mathcal{F}(u_j)$  is a descendant of $\mathcal{F}(u_i) \in T$      
    \item  $\exists$ edge $e: u_i \Rightarrow u_j \in P$ where $label(u_i) = label(\mathcal{F}(u_i))$  and $label(u_j) = label(\mathcal{F}(u_j))$, $ e(c) \leq \mathcal{F}(e(c))$ where $c$ is the frequency count of the relation such as $u_i \Rightarrow u_j$ implies  $e \in R $ 
    \item $ R \subseteq P$

\end{itemize}


\noindent \textit{Scoring Matched patterns:}
While \textit{exact matches} do not require a threshold for detection, \textit{partial matches} require the computation of a score in order to determine if they are anomalous. 
Following pattern matching, we establish a threshold for detecting malicious task trees. That is, pattern task-process-trees with a score greater than the threshold are deemed abnormal.
 
The percentage of items that match is used to calculate the score for a partial match. We consider the same weight or variable weight based on the depth of the element in the tree for the scoring calculation. Our intuition here is to prioritize the matches that are deeper in the tree. For a given pattern $T$, let $R$ be the partial match tree for the pattern, $T$ be the baseline tree, $\omega$ be the weight and $\delta$ represent the threshold. Partial match for the same weight is determined by:

\begin{itemize}
    \item $k= \sum\limits_{i=1}^{|R|} \omega_i (= 1)$  and  $l = \sum\limits_{i=1}^{|T|} \omega_i (= 1)$ 

    \item $x = k/l$
    \item If  $x \geq \delta$  then \textit{Match} Else \textit{Not Match}
\end{itemize}

A partial match with variable weight calculates the pattern matching score based on the item's depth in a tree. Partial match for the variable weight is determined by:

\begin{itemize}
    \item $k= \sum\limits_{i=1}^{|R|} \omega_i (= depth(R_i)) $  and  $l = \sum\limits_{i=1}^{|T|} \omega_i (= depth(R_i)$ 
    \item $x = k/l$
    \item If  $x \geq \delta$  then \textit{Match} Else \textit{Not Match}
\end{itemize}

\subsubsection{Sequential Pattern Analysis.}
\label{Sequential pattern classification}

We propose the use of sequential pattern analysis on the set of task trees as a means for exploiting the temporal nature of the data. Specifically, a set of {\em traces} of activity is extracted from each task tree, where each trace is a sequence of actions performed as part of that task. Given a set of such traces, sequential pattern analysis, using such techniques as sequential pattern mining and classification, can be conducted to identify common patterns of interest, which can then be used to help determine the likelihood of a task containing malicious activity. 

Sequential pattern mining (SPM)~\cite{agrawal1995mining,mooney2013sequential} is a collection of techniques that focus on the identification of frequently occurring patterns of items  (i.e., objects, events, etc.), where ordering of these items is preserved. Let $I$ be a set of {\it items}, and $S$ be a set of {\em input sequences}, where each $s \in S$ consists of an ordered list of {\it itemsets}, or sets of items from $I$, also known as {\it transactions}. A sequence $\left\langle a_1 a_2 \ldots a_n \right\rangle$ is said to be {\it contained} in another sequence $\left\langle b_1 b_2 \ldots b_m \right\rangle$ if there exist integers $i_1, i_2, \ldots, i_n$ with $i_1 < i_2 < \ldots < i_n$ such that $a_1 \subseteq b_{i_1}, a_2 \subseteq b_{i_2}, \ldots, a_n \subseteq b_{i_n}$. A sequence $s \in S$ {\it supports} a sequence $s'$ if $s'$ is contained in $s$. The support $sup(s')$ for a sequence $s'$ given a set $S$ of input sequences is the percentage of sequences in $S$ that support $s'$, and is equal to $sup(s') = \left| \left\{s \in S |s \mbox{ supports } s'\right\} \right|/|S|$. A sequence $s'$ is deemed a {\em sequential pattern} if $sup(s')$ is greater than some pre-specified minimum support. Such a pattern with a total cardinality of its itemsets summing to $n$ is referred to as an {\em $n$-sequence} or {\em $n$-pattern}. A sequential pattern $s'$ is a {\em maximal sequential pattern} in a set $S'$ of sequential patterns if $\forall s'' \in S'$ where $s'' \neq s'$, $s''$ does not contain $s'$. The general goal of sequential pattern mining is then to identify the set $S'$ that contains all (and only those) sequences that are deemed sequential patterns according to the above. In some cases, the set consisting of only maximal sequential patterns is preferred.

Given a supervised learning model in which input sequences are assigned and labeled according to two or more classes, sequence classification~\cite{lesh1999mining,lesh2000scalable,xing2010brief} techniques can be used to attempt to classify new sequences, by using frequent sequential patterns as features in the classification. In addition to the above SPM model, consider the addition of a set of class labels and a labeling function $\ell:S \rightarrow L$ that labels the input set. $S$ is thus a set of {\it examples}, where each example $s \in S$ can be represented by a set of features from the set $SP'$ of frequent sequential patterns. Selected features should exhibit each of the following properties:

\begin{itemize}
\item High frequency
\item Significantly higher representation in one class than the other(s)
\item No redundancy
\end{itemize}

Given these identified features, standard machine learning based classification methods such as SVM or Na\"ive Bayes can be used to build a classification model and label new instances accordingly by considering the feature patterns that they do and do not contain.

Sequential pattern analysis can be used either 1) to obtain a baseline model for normal activity, against which new activity can be measured in order to identify potential anomalies, or 2) to train a supervised classifier on labeled data containing both baseline and malicious activity in order to classify new instances as benign or malicious. In the former case, sequential pattern mining can be used to identify frequent patterns that occur typically in the baseline activity, which can then be used to measure common volumes of noise, i.e. activity that does not conform to identified regular patterns, within the baseline data in order to ascertain a tolerable level. Noise from new activity can then be measured against these tolerance levels, and any excessively noisy activity can then be identified as potentially anomalous. For the latter, a classifier can be trained to detect malicious traces, as outlined above, if the required labeled data exists. 



\section{Experiment \& Evaluation}

Our experiment is conducted on a workstation equipped with an Intel Core i7-10750H processor running at 2.6 GHz, 128 GB of RAM, and six cores. For data extraction, process tree construction, training, and tuning model we use Spark 3.0.2 and Python libraries.

\subsection{Experiment dataset}

We choose OpTC dataset as it enables significant study in the area of process trees. Process trees encapsulate sequential data in log events that are semantically related but chronologically distorted. Enterprise operating systems are, by definition, process-oriented. Each process can be traced back to the process that launched it; each state change can be traced back to the process that caused it to occur.

We evaluate TapTree mainly on OpTC dataset: a benign dataset from 5 hosts, a malicious dataset from 16 hosts. To perform in-depth analysis and mining of log entries within a day/host, we separate each log entry into seven key characteristics (object, action, processID, ParentProcessID, image path, time, and host) and constructed process-trees from the raw events for the benign and malicious hosts. This way, we were able to generate 91,242 process trees from the benign dataset, but only 39 from the malicious dataset. 

We consider user-specific artifacts because the same user may approach a task differently each time it is completed, or because various users may offer similar actions for a given task. Therefore, for benign activity dataset, we target all activities conducted by five targeted hosts over a seven-day period.         

Malicious activities in Darpa-OPTC dataset has been generated mainly from three APT activities on windows systems: Remote Code Execution and Shell Code Injection aka Beacon (Cobalt Strike), Remote Code Execution and Lateral Movement aka Powershell Empire, Credential Harvesting aka Customized Mimikatz. 
APT activities come with an extremely small percentage of any dataset (less than 0.001\% in Darpa-OpTC dataset). For example, Darpa-OpTC's red team targets just 27 hosts out of 1000 networked hosts for launching a series of APT attacks while engaging in benign activities. We capture malicious events from all the hosts over

Prior to matching, the evaluation dataset is filtered to remove trees that do not meet specified thresholds. 
For example, the smallest threshold tested used a minimum of 3 nodes and a minimum depth of 2, resulting in 44608 benign trees and 27 malicious trees, whereas the strictest threshold tested used a minimum of 5 nodes and a minimum depth of 3, resulting in 13,146 benign trees and 16 malicious trees.



\subsection{Evaluation} 


We evaluate the effectiveness of TapTree for behavior abstraction from three aspects.
(1) Comparative study on the effectiveness of tree and sequence based data mining for user behavior footprint (2) How does query accuracy vary when pattern size in queries changes? (3) How does
the amount of training data affect query accuracy? (4) 
The performance penalty associated with various threshold and pattern queries.
Evaluations are conducted on each of the tree pattern queries and sequential pattern analysis approaches using the experiment dataset from DARPA OpTC dataset. 

Tests are carried out at several thresholds such as the percentage of partial match, the minimum number of nodes and depth (tree pattern queries approach), and minimum likelihood of trace maliciousness needed for classification (sequential pattern analysis approach).
Performance is measured using $k$-fold ($k=10$) cross-validation for the larger set of benign instances, and leave-one-out cross-validation for the smaller set containing malicious instances.



To analyze TapTree's accuracy in abstracting host behavior and compare the efficacy of behavior models (see Section \ref{Baseline model comparison}), 
we use five users' data (chosen at random from 1000 windows hosts) to construct a baseline host model, and any user except those five for pattern matching.
Table \ref{Table:Genreation} shows the average run-time required to match a tree. 
While the time required to generate baseline trees from semantic aggregation is longer than that required to generate clustered trees or a Temporal Tree set (4301.83 seconds versus 163.25 and 135.51 seconds, respectively), the result is a significantly smaller number of behavioral trees for matching (36 trees compared to 3501 trees). This reduction in the tree size helps improve tree matching time. That is, compared to Clustered Tree, the average time required to match a new behaviour tree with semantic aggregation can be lowered by 39\%. Although there is a cost associated with the baseline generation time associated with semantic aggregation of trees, the baseline tree generation process only needs to be performed once, whereas matching trees is a recurrent activity.

\begin{figure*}[ht]

\centering

\minipage{0.33\textwidth}
  \includegraphics[width=\linewidth]{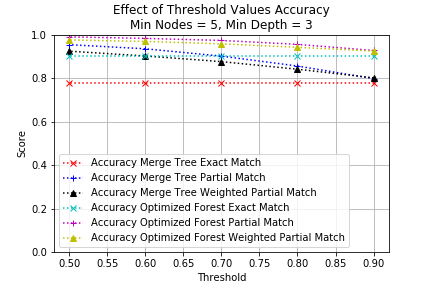}
\endminipage\hfill
\minipage{0.33\textwidth}
  \includegraphics[width=\linewidth]{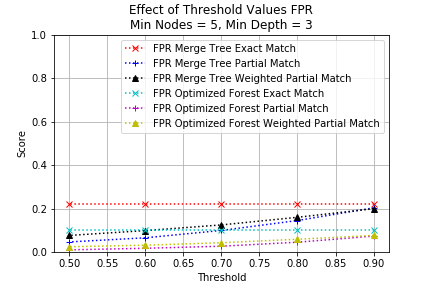}
\endminipage\hfill
\minipage{0.33\textwidth}%
  \includegraphics[width=\linewidth]{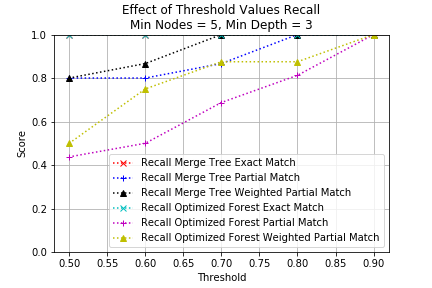}
\endminipage

\caption{Performance of TapTree on different baseline method: Semantic aggregation vs clustered Trees.}
\label{figOFThresholds-a}

\end{figure*}

\begin{figure*}[ht]

\centering


\minipage{0.33\textwidth}
\includegraphics[width=\linewidth]{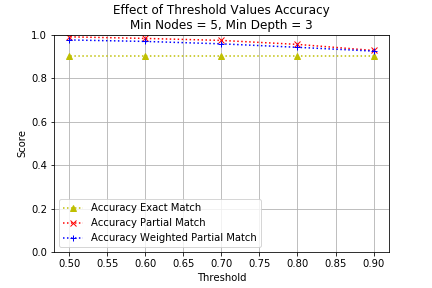}
\endminipage\hfill
\minipage{0.33\textwidth}
  \includegraphics[width=\linewidth]{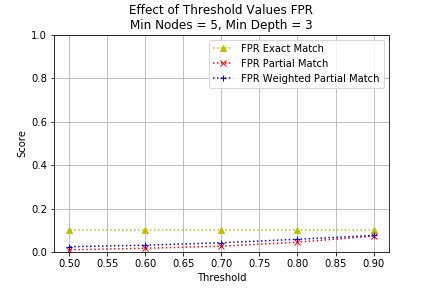}
\endminipage\hfill
\minipage{0.33\textwidth}%
  \includegraphics[width=\linewidth]{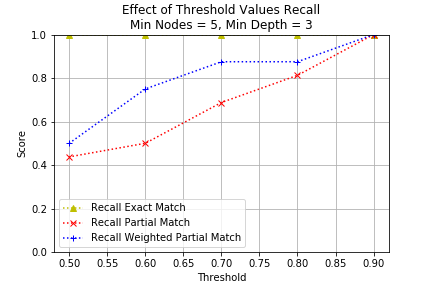}
\endminipage
\caption{Performance of TapTree on different threshold (baseline method: Semantic aggregation)}
\label{figOFThresholds-b}

\end{figure*}

\begin{table}[ht]
\centering
\caption{TapTree performance compared to other Methods on OpTC dataset \cite{OpTC2020}. TapTree data is in the \textit{tree} format (\#Minimum depth = 2, \#Minimum nodes = 3)}

\begin{tabular}{|c|cc|}
\hline
\textbf{Method}           & \textbf{Accuracy} & \textbf{FP Rate} \\ \hline
TapTree (Semantic aggregation) & 0.9913   & 0.008   \\
TapTree (clustered trees)       & 0.9739   & 0.026  \\
TapTree (sequence mining)       & 0.9901   & 0.001  \\
DeepTaskAPT(Trace) \cite{Mamun2021}         & 0.9854   & 0.011   \\
DeepTaskAPT(Task)  \cite{Mamun2021}        & 0..9641   & 0.006  \\
DeepLog \cite{Du2017}         & 0.8354   & 0.161   \\
Random Forest    & 0.9052   & 0.083  \\ \hline
\end{tabular}
\subcaption{partial match threshold = 0.5}
\label{Table:Compare Accuracy}
\bigskip

\begin{tabular}{|cc|}
\hline
\textbf{Method}      &  \textbf{Recall} \\ \hline
TapTree (Semantic aggregation)    & 1.0 (FPR $\approx 0.029 $ )   \\
TapTree (clustered trees) &    1.0   (FPR $\approx 0.067 $ )   \\
TapTree (sequence mining) &    0.8571      \\
DeepTaskAPT (Trace)  \cite{Mamun2021}   & 0.7587 \\
DeepTaskAPT (Task)   \cite{Mamun2021}   & 0.8299 \\
DeepLog    \cite{Du2017}  & 0.7202 \\
Random Forest    & 0.6784  \\ \hline

\end{tabular}
\subcaption{partial match threshold = 0.9}
\label{Table:Compare Recall}

\end{table}

\noindent {\bf Tree Pattern Matching Methods:}
 TapTree methods were found to perform extremely well when compared with existing methods, particularly when leveraging various threshold scores for partial matches as well as conducting proper tree filtering based on minimum node and depth values. Tables \ref{Table:Compare Accuracy} and \ref{Table:Compare Recall} depict results of TapTree methods (Semantic aggregation, Clustered Tress and Sequence Mining) against existing approaches (DeepTaskAPT (Trace)  \cite{Mamun2021}, DeepTaskAPT (Task)  \cite{Mamun2021}, DeepLog    \cite{Du2017} and Random Forest). When using weighted partial matching filtering based on five \textit{minimum nodes} and three \textit{minimum depth} for trees, Semantic aggregation achieved high accuracy (0.9913) over all existing approaches, while posting a false positive rate just slightly higher than DeepTaskAPT (Task). Increasing the partial match percentage threshold score from 0.5 to 0.9 allows both TapTree methods to achieve a recall score of 1.0, which means that they were able to capture all malicious tasks, while only increasing false positive rates to 0.029 and 0.067, respectively. While the Sequence Mining TapTree method posted accuracy and recall scores slightly lower than semantic aggregation, it produced the lowest false positive rate of all methods, and outperformed all existing methods for each of the accuracy, fp-rate and recall metrics. Performance of sequential pattern-based analysis is examined in detail in the next section.

\begin{table}[ht]
\centering
\caption{TapTree performance with different Threshold scores and tree pattern queries algorithms (baseline model: semantic aggregation, Min Node = 5, Min Depth = 3) }
\begin{tabular}{|ccccc|}
\hline
 Method  & Threshold & Accuracy & Recall & FP Rate   \\ \hline
          Exact Match &  --  & 0.900547 & 1.0       & 0.099574 \\ \hline
     & 0.9   & 0.926911 & 1.0       & 0.073178 \\
 Partial Match      & 0.7 & 0.973028 & 0.6875  & 0.026624 \\
 (same weight)      &  0.5 & 0.989439 & 0.4375  & 0.009889 \\ \hline
       & 0.9    & 0.923948 & 1.0       & 0.076145 \\
 Partial Match      &  0.7   & 0.957301 & 0.875   & 0.042599 \\
(variable weight)        &  0.5  & 0.975232 & 0.5     & 0.02419  \\ \hline
\end{tabular}
\label{Table:Thresholds}

\end{table}

\begin{figure*}[ht]

\caption{TapTree performance when jointly tuning minimum number of nodes and tree depth (N-D) }
\label{figMTFilter}
\centering
\minipage{0.33\textwidth}
  \includegraphics[width=\linewidth]{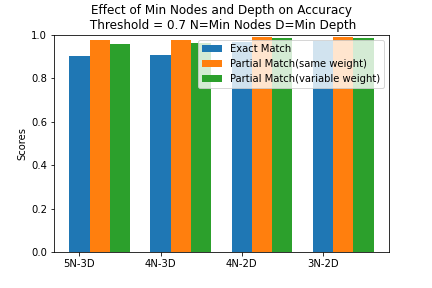}
\endminipage\hfill
\minipage{0.33\textwidth}
  \includegraphics[width=\linewidth]{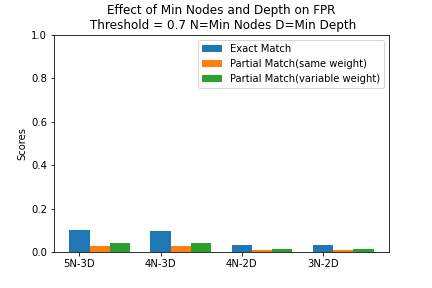}
\endminipage\hfill
\minipage{0.33\textwidth}%
  \includegraphics[width=\linewidth]{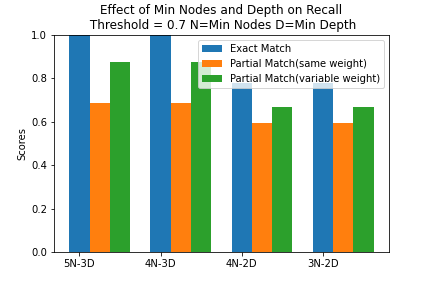}
\endminipage
\subcaption{semantic aggregation}

\centering
\minipage{0.33\textwidth}
  \includegraphics[width=\linewidth]{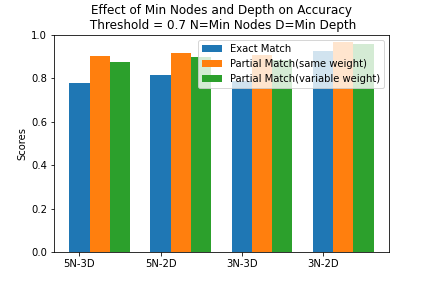}
\endminipage\hfill
\minipage{0.33\textwidth}
  \includegraphics[width=\linewidth]{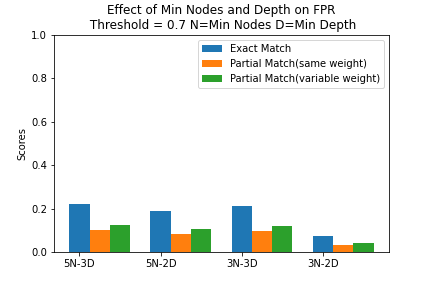}
\endminipage\hfill
\minipage{0.33\textwidth}%
  \includegraphics[width=\linewidth]{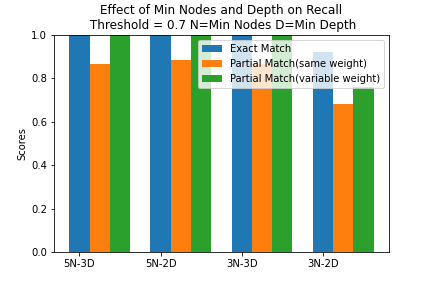}
\endminipage
\subcaption{Clustered trees}

\end{figure*}

Table \ref{Table:Thresholds} illustrates the performance of tree pattern queries algorithms discussed in Section \ref{Tree pattern queries} based on semantic aggregation baseline model (see Section \ref{Baseline model}).
Except in cases when perfect matches are required, matching threshold scores (discussed in section {Tree pattern queries}) has a significant impact on TapTree's anomaly detection performance. 
The use of partial matches enables finer tuning of TapTree's performance to optimize results for recall, accuracy, and false positive rate (FPR). A higher threshold score value ensures improved recall performance, whereas a lower score improves accuracy and FPR. 

Figure \ref{figOFThresholds-a} depicts the comparative performance on semantic aggregation vs clustered tree with respect to accuracy and recall. As the semantic aggregation baseline method achieves the best performance, Figure \ref{figOFThresholds-b} describes the impact of partial match thresholds with a given minimum number of nodes and tree depth. In Figure~\ref{figMTFilter}, we jointly tune \textit{minimum number of nodes} (N) and \textit{tree depth} (D). TapTree clearly achieves the best performance when N-D is 3-2 (accuracy/FPR) and 5-3 (recall) for a given threshold of 0.7. \\

\begin{table}[h]
\centering
\caption{Results of sequential pattern-based malicious behaviour detection, at various classification thresholds}
\begin{tabular}{|c|c|c|c|c|c|} \hline
Threshold & TPR & TNR & Precision & Accuracy & FPR \\  \hline
0.1 & 1        &   0       &   0.00037 & 0.00037 & 1 \\
0.2 & 1        &   0       &   0.00037 & 0.00037 & 1 \\
0.3 & 0.92063  &   0.54612 &   0.00074 & 0.54633 & 0.45388 \\
0.4 & 0.86243  &   0.99012 &   0.03080 & 0.99021 & 0.00988 \\
0.5 & 0.83069  &   0.99717 &   0.09653 & 0.99724 & 0.00283 \\
0.6 & 0.67196  &   0.99936 &   0.27632 & 0.99938 & 0.00064 \\
0.7 & 0.42857  &   0.99997 &   0.81375 & 0.99990 & 0.00003 \\
0.8 & 0.30159  &   1       &   1       & 0.99988 & 0 \\
0.9 & 0        &   1       &   -       & 0.99978 & 0 \\ \hline
\end{tabular}
\label{table:SPM_results}
\end{table}

\begin{figure}[h]
\begin{center}
\pgfplotsset{
small,width=6.75cm,
}

\begin{tabular}{cc}

\begin{tikzpicture}
\begin{axis}[
    title={ROC Curve},
    xlabel={False positive rate},
    ylabel={True positive rate},
    xmin=0, xmax=1,
    ymin=0, ymax=1,
    xtick={0,0.2,0.4,0.6,0.8,1},
    ytick={0,0.2,0.4,0.6,0.8,1},
    ymajorgrids=false,
]
 
\addplot[
    color=black,
    mark=dot,
    ]
    coordinates {
		(1,1)(1,1)(0.45388,0.915343915)(0.00988,0.857142857)(0.00283,0.825396825)(0.00064,0.666666667)(0.00003,0.428571429)(0,0.301587302)(0,0)
    };
 
\addplot[
    color=black,
    mark=*,dashed,
    ]
    coordinates {
		(0,0)(1,1)
    };

\end{axis}
\end{tikzpicture}

\end{tabular}
\end{center}
\caption{Receiver operating characteristics (ROC) curve to demonstrate performance of sequential pattern-based classifier at various threshold values}
\label{fig:SPM_analysis_ROC_curve}
\end{figure}
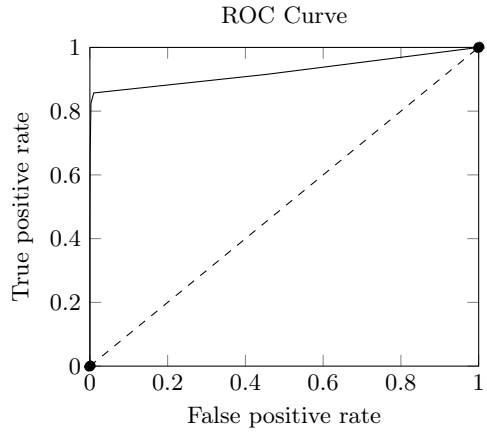

\noindent {\bf Sequential Pattern-Based Methods:}
Performance was also measured for a sequential pattern-based classifier tasked with discerning malicious activity from benign, trained on labeled benign and malicious data extracted from the task trees, using techniques introduced in section \ref{Sequential pattern classification}. In all, the training set consisted of 5,15,888 benign traces and 189 malicious cases, with $k$-fold ($k = 10$) cross-validation employed for the benign cases, and leave-one-out cross-validation employed for the malicious cases. Testing was conducted at various thresholds, where for a threshold of $x$, a test case was classified as malicious if and only if the classifier's deemed likelihood of maliciousness was greater than or equal to $x$.

Table~\ref{table:SPM_results} shows performance at each threshold, in terms of true positive rate TPR (i.e. recall, the percentage of malicious cases correctly classified as such), true negative rate TNR (i.e. the percentage of benign cases correctly classified as such), precision, accuracy and false positive rate FPR. Figure~\ref{fig:SPM_analysis_ROC_curve} depicts the receiver operating characteristics (ROC) curve, plotting the true positive rate against the false negative rate.

Results show that the method is highly effective at correctly classifying both benign and malicious traces, as evidenced by the high true positive and negative rates, particularly at the 0.4 and 0.5 thresholds, as well as an ROC curve that lies well to the top and left  of the dashed line in the graph representing random guess.

Precision is low at most levels due to the highly unbalanced nature of the data, meaning that even high accuracy levels can result in a large number of false positives. For example, at the 0.5 threshold, benign cases are correctly classified almost 98\% of the time, however this amounts to 1460 false positives compared to 156 true positives at that threshold level. However, at higher threshold levels the precision performs remarkably well, reaching 81\% at 0.7 with only 16 false positives (FPR = 0.0003) while still identifying 43\% of malicious cases, and 100\% precision at 0.8 while still identifying 30\% of malicious cases.



\section{Related work}
Sequence approaches such as \cite{LogNADS2021,Du2017,Shen2018,APTmodel2,Mamun2021} take log entries and concatenate them chronologically into sequences. These techniques are primarily concerned with capturing temporal and sequential connections between log entries, and often make use of deep learning techniques such as Long Short-Term Memory (LSTM) or machine learning tool such as signature kernel, to learn from previous events and forecast future events. Although deep learning, like LSTM, may recall long-term dependencies in sequences, it does not compare every behavior of the user explicitly \cite{log2vec2019}, and ignores interactive relationships between events or hosts \cite{log2vec2019,Mamun2021}. This can hinder performance possibly prevent effective identification of APTs. Additionally, some of them demand a considerable amount of labeled (malicious) data during the training process or a high number of features for model creation that might not be available in real-world deployment. In \cite{Mamun2021,log2vec2019}, the authors address some of these issues through alternative methods such as finding logical relationships between user tasks prior to applying deep learning \cite{Mamun2021}, and utilizing a graph that depicts a user's interaction with hosts \cite{log2vec2019,graph2021}.

Meanwhile, a recent approach known as {\em SK-Tree} \cite{cochrane2021sk} uses streaming trees to represent computer processes, and presents a malware detection algorithm leveraging a machine learning tool (signature kernel \cite{signatureKernal}) for time series data, with promising results. While the SK-Tree study focused on one day of data from a single user (0201), we attempt to expand our reach and leverage more of the dataset to include data captured from multiple users over multiple days.


\section{Limitations}
This section discusses some of the inherent limitations of the design choice, as well as the ramifications and potential extensions of this work.

TapTree's design relies on process-tree to abstract host activities. Therefore, it may not be effective at detecting attacks that do not result in spawning new processes in the operating system. For example, attacks such as buffer overflows, which do not involve the creation of a new process, are not protected by TapTree. Baseline model may require periodic retraining due to semantic shifts in user/host behaviour and addition of previously unseen behaviour patterns. An analyst can identify new host behaviours over the course of time, sanitise them, and decide carefully whether to include the new behaviours in the baseline model for re-training.
Our empirical experiment shows that TapTree can recognise an unknown tree pattern in milliseconds, while retraining with fewer new patterns takes seconds.

The lack of attack dataset on which to train malicious behavior classifiers prompts further investigation into sequential pattern-based methods for developing baseline models.
Also worthwhile of further study is a comparison of our methods to relative graph-based schemes where scalability is an issue, as well as investigation into more efficient partial matching, a robust baseline model, and the exploration of new tree matching algorithms.

Despite the algorithms' poor worst-case time complexity for constructing the initial baseline tree, which is quadratic in the number of inputs, experimental evaluations show that they perform exceptionally well in terms of average run time, especially for pattern matching.

\section{Conclusion}

We present a detailed study on the effectiveness of performing sequential pattern matching for anomaly detection. To facilitate this, we present TapTree, a task-process-tree based model for APT detection on system log data that represents host log data in such a way that facilitates detection of malicious behaviour. Two distinct approaches for this detection are explored. The first attempts to match new data to existing baseline behaviour, represented as \textit{temporal trees}, in an attempt to identify anomalies that could signify attack behaviour. The second extracts sequential behaviour called \textit{traces from the trees} for existing baseline and malicious behaviour samples, and uses sequential pattern mining to identify critical patterns for use in a malicious behaviour classification model.

As for detection performance evaluated using the DARPA OpTC dataset, we demonstrate that one particular TapTree tree matching algorithm, semantic aggregation, achieved high accuracy over all existing approaches, while both semantic aggregation and Clustered Trees were found to achieve a perfect recall by adjusting tree matching thresholds, while still maintaining low false positive rates. The sequential pattern-based TapTree method, on the other hand, posted the lowest false positive rate of all methods, and outperformed all existing methods for each of the accuracy, fp-rate and recall metrics.


\section{Acknowledgement}

We would like to thank the Communications Security Establishment Canada team, especially Dr. Benoit Hamelin for supporting the project and providing the materials needed for this work. A special thanks to Kevin Shi from the University of Windsor for all the support during his co-op term with NRC.

%
%
%
%
\bibliographystyle{unsrt}

\bibliography{sample-base.bib}

\end{document}